\documentclass[aps, prd, twocolumn, lengthcheck, superscriptaddress, % showpacs,
nofootinbib]{revtex4-1}

\usepackage{epsfig}
\usepackage[usenames]{color}
\usepackage{graphicx}
\usepackage{amsmath}
\usepackage{epstopdf}

\def\bi{\bibitem}

\def\la{\langle}\def\ra{\rangle}
\def\be{\begin{eqnarray}}\def\ee{\end{eqnarray}}
\def\lsim{\mathrel{\rlap{\lower3pt\hbox{\hskip1pt$\sim$}}
     \raise1pt\hbox{$<$}}} %less than or approx. symbol
\def\gsim{\mathrel{\rlap{\lower3pt\hbox{\hskip1pt$\sim$}}
     \raise1pt\hbox{$>$}}} %greater than or approx. symbol

\def\CHIEFTpi{$\chi$EFT$_\pi$}
\allowdisplaybreaks

\begin{document}

\title{A Bottom-Up EFT  Approach To  Superdense Baryonic Matter}

\author{Mannque Rho}
\email{mannque.rho@ipht.fr}
\affiliation{Universit\'e Paris-Saclay, CNRS, CEA, Institut de Physique Th\'eorique, 91191, Gif-sur-Yvette, France }

\date{\today}

\begin{abstract}
 
 How to arrive at the densest matter in massive compact stars  starting from Walecka's linear $\omega$-$\sigma$ mean-field model is described in a series of arguments anchored on hidden local symmetry, hidden scale symmetry and emergent parity-doublet symmetry. I follow the bottom-up approach from chiral symmetry with pions, coupled to  hidden local and scale symmetry degrees of freedom.  Exploiting the renormalization-group treatment \`a la Shankar and Polchinski of the fermionic interactions on the Fermi sphere, leading to Landau-Migdal Fermi-liquid, one obtains a sort of  ``generalized Covariant Density Functional" that allows via a topology change (quasi-)hadrons to transform to (quasi-)quarks  without phase changes at the center of massive stars.  The highly dense matter  is ``pseudo-conformal" with the sound velocity $v_{pcs}^2/c^2\approx 1/3$ but the trace of the energy-momentum tensor is not equal to zero, hence the matter is non-conformal.

\end{abstract}

%\CHIEFTpi "\def\CHIEFTpi{$\chi$EFT$_\pi$}"

\maketitle

In accessing the extreme state expected to be encountered at high density $n\gsim 3 n_0$  (where $n_0\simeq 0.16$fm$^{-3}$) in massive compact stars, the appropriate theory, assuming gravity is taken to be fully under control, is considered to be QCD .  Up to date, there are no model-independent  approaches. The only tools currently available are effective field theories (EFTs ), one from bottom-up density  n density and the other top-down from asymptotic density. The non-perturbative approach on lattice in QCD is lacking in the density range involved. One can access strongly interacting matter reliably from near zero density to near the nuclear matter equilibrium density $n_0$ in the approach nowadays known as chiral effective field theory (EFT).  This approach  denoted  as ``\CHIEFTpi" with the nucleons  interacting with nearly zero-mass pions $\pi$ -- and with no other degrees freedom -- was aptly phrased by Weinberg in his ``Folk Theorem on EFT"~\cite{FT}.  It will be referred to as FT. (At very low energies, even the pions can be ``integrated out" giving rise to what's known as ``pionless EFT," nicely exemplifying the notion of the Folk Theorem.)  This theory is found to be highly successful for  low-energy nuclear interactions, thus reliably describing nuclear matter up to, say,  around $\sim n_0$. Applied to finite nuclei as well as to infinite nuclear matter, the \CHIEFTpi  is giving a fairly successful description of finite nuclei as well the EoS of infinite nuclear matter, and the results are often cited as a ``first-principles" rendering of nuclear physics. However as an EFT in terms of a given power counting, it is bound to ``break down" at a scale of momentum or density not far above $\sim n_0$. An equally serious conundrum  is bound to occur at the same range of density coming top-down in density from the asymptotic regime. Thus the center of massive compact stars remains a totally unexplored area.

In this note, I wish to summarize  in the simplest terms feasible, leaving out non-crucial details,  what is in the bottom-up scheme that  offers a possible valid alternative to \CHIEFTpi with the possibility to by-pass the stumbling block to the higher density regime involved. For reasons that will be made clear,  it is found to be more appropriate  to adopt an effective field theory in spirit paralleling the \CHIEFTpi  strategy \`a la FT instead of in widely employed patch-work models built on incomplete (experimentally measured)  data and phenomenological models developed.  Given the necessary presence of (pseudo-Goldstone) pions $\pi$ in nuclear dynamics,  it will be found to be astute to adopt the bulk baryonic matter to be a Fermi liquid in the presence of  pion cloud, namely ``Landau-Migdal Fermi liquid" with the pion field  included.

In its present form,   the arguments are often too short of sufficiently satisfactory rigor,  but I will indicate how to make some of the reasonings more acceptable.  I opt to cite in the bibliography only a few that are crucial to the matters discussed,  putting what's considered to be relevant but not essential for the logic in the footnotes.
 
 The basic initial idea of formulating what  I will call ``G$n$EFT" for dense matter that follows originates  from $\sim$ 1974 with Walecka's linear mean-field model~\cite{walecka}.   Among the various properties of the model assumed, some of which turn out to be more or less  invalid from the present-day's understanding, what turned out to be highly relevant for dense compact star matter which was stressed in \cite{walecka} -- and will be reiterated --  was that the relativistic mean-field approximation of the linear model became better justified at higher baryon densities. The degrees of freedom taken first were the $\omega$ and the scalar $\sigma$ fields coupled to the nucleons, the former accounting for  simulating strong repulsion and the latter for the compensating attraction necessary for stable baryonic matter.  I must stress here that this interplay between the $\omega$ and  $\sigma$ --  together in the presence of other additional quantum-number degrees of freedom -- turns out to play the crucial role forming the baryonic matter in a strong-coupled Fermi-liquid structure on crystal lattice -- that leads to  compact-star matter at high density~\cite{matsui}.   
 
 The interplay of the vector $\omega$ and the scalar $\sigma$ -- which will be put below in the form of the ``conformal compensator field" ${\chi}=f_\chi e^{\ \sigma_d/f_\chi}$  replacing the scalar $\sigma$ by $\sigma_d$  with $d$ standing for ``dilaton" in what follows -- plays the most important role in the development of the arguments that lead to the formulation of the theory G$n$EFT. It provides the key ingredient in the present formulation of the notion of ``pseudo-conformality" that governs the properties of both normal and and superdense matter. Not widely recognized in nuclear physics circle is that Walecka's mean-field structure of the ground state of multi-baryon systems can actually be obtained from Weinberg's Folk Theorem. The NLO (next-to-leading order)  Weinberg's nuclear chiral Lagrangian \CHIEFTpi when expressed in terms of four-nucleon terms at the mean field approximation gives essentially the same result as Walecka's linear mean field as density increases. This indicated that the Walecka model captures a lot more than just the linear $\omega$-$\sigma$ coupled model but a part of a more {\it fundamental} theory embodying chiral symmetry as pointed out by particle theorists~\cite{gelmini-ritzi} and  exploited in nuclear theory~\cite{DD}.  This then makes Walecka's approach fully consistent with the {\it modern notion} of the ``renormalizability" \`a la FT. There was no reason to restrict to only the $\omega$-$\sigma$ mesons. Massive degrees of freedom embodying e.g. ``hidden local symmetric" (HLS) vectors, ``hidden scalar" (HS) mesons, i.e., dilatons, infinite towers of heavy mesons such as ``holographic" QCD fields etc. can naturally figure to account for not only charge-symmetric nuclear matter ground state but also various quantum-number excitations leading to what  resembles -- and generalizes -- the  ``Density Functional Theory"  \`a la Hohenberg-Kohn theorem developed in condensed matter physics. Let me denote in what follows by $\cal{L}_{\rm GnEFT}$ the Lagrangian so constructed with  the relevant  fields  and their symmetries duly accounted for to lead to the theory referred to as G$n$EFT.  
 %\CHIEFTpi%%%%
%%% L_{\rm GnEFT}%%%

 Aside from baryons (i.e., nucleons and strange baryons) and pseudo-Goldstone pions,  $\cal{L}_{\rm GnEFT}$  will comprise  of the low-lying vector mesons $\omega$ and $\rho$ belonging  to  ``hidden local symmetry (HLS)" mesons referred to  in the review \cite{HLS}.\footnote{Generally the established low-energy theorems such as KSRF etc. do properly hold in this version of HLS except for the Vector Dominance  for the photon coupling to nucleons where the infinite tower of vector mesons is required as seen in the Sakai-Sugimoto holographic QCD model as pointed out in \cite{hryy}. This is an aspect of Cheshire Cat Principle~\cite{CC} holding via hidden symmetries  in the working of G$n$EFT.}  And of course a scalar degree of freedom to counter-balance the $\omega$ repulsion will be absolutely needed as in the Walecka model. Note that it is {\bf not} the scalar $\sigma$ as in the linear $\sigma$ model of chiral symmetry. It is actually not visible in the particle data, so it is properly associated with a ``hidden symmetric" scalar comparable in mass to the $\omega$. It will turn out to figure in  $\cal{L}_{\rm GnEFT}$ as a pseudo-dilaton, a Goldstone boson of the broken scale symmetry, with a mass $\sim 600$ MeV in free space. 

How the scalar dilaton figures in QCD has a long history, remaining still an unresolved issue. It cannot be adequately explained in a short note as this. It enters at the scale going beyond QCD, such as a ``conformal Higgs" boson,  in cosmology etc. For the problem at issue,  the superdense matter,  involving $N_f\lsim 3$ for the number of flavors, this un-resolvability turns out not to be too serious. It can be suitably finessed.  The approach adopted in G$n$EFT relies on the recently discussed notion of ``Genuine Dilaton (GD)"~\cite{GD} and ``QCD-conformal dilaton (QCD-CD)"~\cite{zwicky} which posit the presence of an Infra-red fixed point (IRFP) with vanishing  $\beta_{IRs}=0$ in QCD for low $N_f$, $N_f \lsim 3$.  {\it The distinguishing feature of the GD/QCD-CD is that at the IRFP, the dilaton decay constant $f_{\sigma_d}$ is not zero although the trace of the energy momentum tensor $\theta_\mu^\mu$ goes to zero.   Matter fields such as $N$, HLS mesons etc. remain massive at the fixed point.  In fact the NG mode for conformal invariance is compatible with the NG mode for chiral invariance.}

One can now proceed to look into the structure of G$n$EFT built on the Lagrangian $\cal{L}_{\rm GnEFT}$. One writes the Lagrangian  schematically as
\be
\cal{L}_{\rm GnEFT}= \cal{L}^{\rm{Sinv}} (\rm{N}, {\rm HLS}, \pi, \chi_{\rm d}) + V(\chi_{\rm d}, {\rm matters})\label{Lag}
\ee
where $\cal{L}^{\rm{Sinv}}$, the action of which  consist of the terms made scale-invariant by multiplying by powers of the conformal compensator fields on the matter fields, say, nucleon, HLS as well as HS fields,  and $V$ is the scale-symmetry breaking potential term including the trace anomaly as well as mass terms. How the potential $V$ is written differs between the GD and QCD-CD approaches but the predictions for compact-star observables turn out come out to be  the same to the (chiral-scale) order involved.  

As stressed, this Lagrangian is constructed, in additions to the ramifications mentioned, with one specific feature that is not implemented in effective theories \`a la FT with the pion fields only. It has to do with a mechanism to allow the hadron-quark (HQ) changeover/continuity -- with or without phase changes -- that is widely considered to be taking place in compact stars at densities exceeding $\sim (2-3)n_0$. 

Now suitably implemented with heavier vector mesons,    $\cal{L}_{\rm GnEFT}$, leads, in consistency with the notion of FT, to what can be considered as a ``covariant density functional theory (CDFT)" endowed with, in addition, Fermi liquid properties~\cite{MR}. This sets up what I call the Landau(-Migdal) Fermi liquid basis.\footnote{From here on Landau Fermi liquid, relevant for dense matter, is to imply Landau-Migdal Fermi liquid.}.   This series of arguments anchored on a ``sliding-vacuum"  structure of baryonic (BR-scaled) matter captured by the vacuum condensate $\la\chi_d\ra$ (which is not necessarily equal to $\la\bar{q}q\ra$) plays an important  role which is not adequately taken into account in many of the density-functional approaches found in the literature. It  is essentially captured in certain topological property of the matter in the ``double-decimation (DD)" idea reviewed in \cite{DD}.  Now the most {\it crucial observation} to make here is that the Walecka mean-field approximation of the linear $\omega$-$\sigma$ model corresponds to the core part of the Fermi-liquid fixed-point (FLFP) approximation in the Renormalization-Group (RG) treatment of the effective theory of interacting fermions on the fermi surface~\cite{shankar,polchinski} with the Lagrangian (\ref{Lag}) with the parameters  {\it fixed} at the given density. The defect of too large an incompression modulus found in the linear $\omega$-$\sigma$ model is remedied by the hidden degrees of freedom included in  the Lagrangian ${\cal{L}_{\rm GnEFT}}$. As an EFT, the power corrections are taken over by $1/\bar{N}$ corrections where $\bar{N}=(\Lambda_{F}-k_F)/k_F$ with  $\Lambda_{F}$ the cutoff measured relative to the origin of the Fermi sphere.

% $\cal{L}_{\rm GnEFT}$
 
How this approach on the Fermi surface works was  worked out in \cite{FR}, motivated by the Shankar/Polchinski work~\cite{shankar,polchinski}.\footnote{There nuclear low-energy theorems were verified to be very well satisfied. 

(A) Nuclear electromagnetic response functions at $q\to 0, \omega\to 0$ limit predict
\be
\delta g_l=\frac 49 [1/\Phi -1-\frac 12 \tilde{F}_1^\pi]\tau_3\nonumber
\ee 
where $\tilde{F}_1^\pi$ is the Landau fixed-point  interaction term coming from the pion exchange and $\Phi$ is the BR scaling accounting for how the diilaton decay  constant scales in baryonic matter.  In $^{208}$Pb, this gives $\delta^{\rm proton} g_l= 0.22$ which agrees with the data $\delta g_l^{\rm proton}=0.23\pm 0/03$. 

(B) What's predicted in the weak axial current is  the Landau fixed-point $g_A^L$ given by the Goldberger-Treiman relation at large $N_c$ limit
\be
g_A^L=g_A (1-\frac 13 F_1^\pi)\nonumber
\ee which gives at nuclear matter density 
\be
g_A^L\approx 1.\nonumber
\ee
This was interpreted as the effective $g^{\rm eff}_A\approx 1$ in heavy nuclei, e.g.,  $^{100}$Sn Gamow-Teller transition. There is a disagreement in the experiments by the two Labs, GSI and RIKEN which might indicate a fundamental effect -- due to an anomaly -- in the effective constant in $g_A$, which calls for confirmation for the 2$\nu\nu$ beta decay process for probing BSM. This issue was discussed in detail in  \cite{2-neutrino}.}

This then suggested it logical to generalize the Walecka model  exploiting the EFT Lagrangian  $\cal{L}_{\rm GnEFT}$  for nuclear processes in complex finite nuclei and nuclear matter up to density not just to $\sim n_0$ but also higher.  In other words, this  could provide the derivation of  a ``Generalized Covariant Density Functional (GCDF) Theory" with the parameters of the Lagrangian fixed by G$n$EFT, not by  arbitrary fitting.\footnote{I should mention here for those who favor an axiomatic approach that there was  an attempt to turn a CDF approach into a formal field theory for particle theory, perhaps in view of string theory in development. See e.g. T. Banks, ``Density Functional Theory for Field Theorists I," e-print 1503.02925.  But I am not aware of any further developments in that line. It would be a challenge to use that approach to the physics of superdense astrophysical matter.}   In fact, there is in the literature  an extensive effort to construct phenomenological CDFs to go even beyond  $\gsim 2n_0$  to confront  compact-star matter. A recent highly sophisticated analysis along this line~\cite{holt} seems to support this feasibility. Up to $\sim 2 n_0$, the approach seems to face  no serious difficulties.  However going towards $\sim (6-7) n_0$ in the center of massive stars must raise a serious question:  How can the putative hadron-quark (HQ) changeover, either continuous or  with phase transition(s), be manifested in the parameterization of the density functional? There cannot be a clear answer to this question, given the absence of model-independent tools -- such as lattice QCD -- at high density.

It is here that topology enters in G$n$EFT~\cite{PKLMR,MaRho-Rev}.\footnote{In these reviews, much of the inaccurate numerics and erroneous statements have been corrected from the original version \cite{PKLMR} which contains essentially all the details of what's found/corrected in these review articles.}

It has been argued that at large $N_c$, baryons on crystal lattice, described in terms of skyrmions,  fractionalize into half-skyrmions with no changes in the chiral condensate, hence involving no phase changes\footnote{See, e.g., {\it The Mulitfaceted Skyrmion} \ (World Scientific 2017) Ed. by Mannque Rho and Ismail Zahed. Unfortunately due to mathematical difficulties this argument has been given strong support up to date  neither by mathematicians nor by theoretical physicists, this even for the Skyrme's original model.}. The skyrmion-to-1/2 skyrmion changeover density $n_{1/2}$ is not fixed by theory, but it seems natural to assume $n_{HQ} \sim n_{1/2}\sim (2-3)n_0$.  The FLFP approximation with corrections going like $1/\bar{N}\sim 1/k_F$, interpreted in the half-skyrmion lattice system, gives predictions that must be different from the phenological CDF results. 

Now how different can they be?

First to note is that at the Fermi-liquid fixed-point (FLFP) (or mean-field) approximation, i.e., $1/\bar{N}\to 0$, nearly all nuclear properties  {\it predicted parameter-free} with the hidden symmetry degrees of freedom taken into account  for densities $\lsim 2.5 n_0$ are more or less all compatible with the phenomenological CDF results although  there can be $O(1/\bar{N})$ corrections to the FLFP approximations which of course may be needed for more precise predictions.\footnote{I will comment below how to make such corrections following current developments in condensed matter physics.} 

%%%%%%% to be continued %%%%%%%%%
%%%%%%%%%%%%%%%%%%%%%%%

However with the parameter changes incurred by the skyrmion-to-1/2 skyrmion crossover at $n_{1/2}$, one expects -- and indeed finds -- certain  dramatic modifications in the   EoS as the density exceeds  $n_{1/2}$.    Indeed among various changes,   one striking modification is the sound velocity in the center of massive stars.
\begin{figure}[h]
\begin{center}
\includegraphics[width=4.9cm]{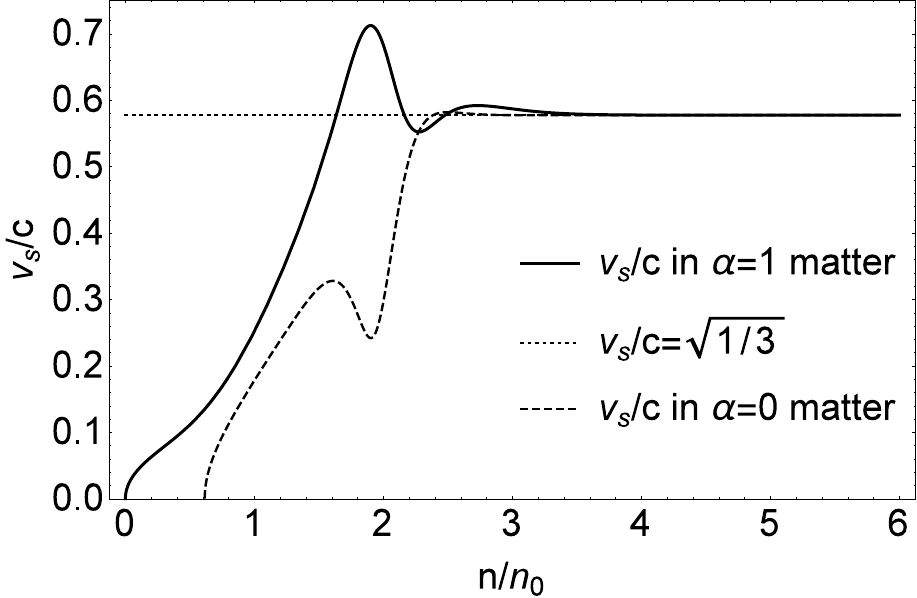}
\includegraphics[width=6.7cm]{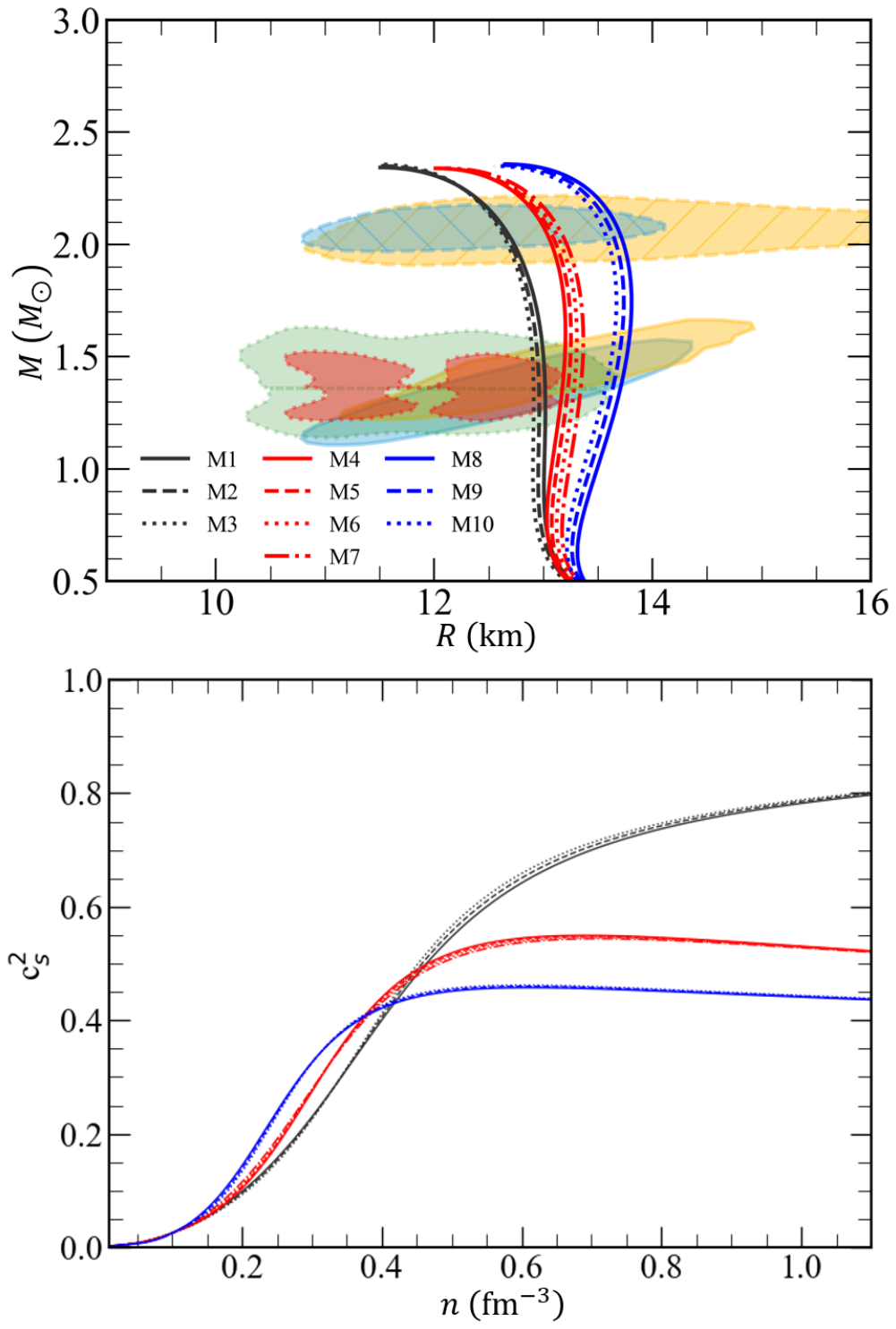}
\caption{Top: Sound velocity $v_{pcs}$ predicted in G$n$EFT for $n_{HQ}=2.5 n_0$.  Middle \& Bottom:   Sound velocities (bottom) predicted in an  RMF calculation~\cite{holt} with  $M(M_\odot))$  vs. $R$ as inputs.} 
\label{sound}
\end{center}
\end{figure}
%
 %One of the most striking phenomena  is the sound velocity in the center of massive compact stars predicted in G$n$EFT for $n_{1/2}\approx 2.5n_0$. 
This is seen in the sound speed predicted in G$n$EFT shown in Fig.~\ref{sound} (top).  Affected with the $\rho$ decoupling $g_\rho\to 0$ and the dilaton $\chi_d$ moving toward   the ``dilaton-limit fixed point (DLFP)"  which ultimately goes toward the limit $\la\chi_d\ra\propto {\rm const.} \to 0$,  the EoS gets  constrained to $g_A\to 1$, $ m_N^\ast \propto f_{\chi_d}\to f_\pi\propto m_0$ where $m_0$ is the chiral-invariant mass arising from the parity-doublet symmetry etc.,  and  the sound speed $v_{pcs}$  (where the subscript pcs corresponds to ``pseudo-conformal sound speed")  becomes extremely simple as the density goes over  $n_{1/2}$.  This property emerges at the Fermi-liquid fixed point  with all the component fields becoming  non-interacting scale-invariant  quasiparticles.\footnote{This feature can be explicitly seen in the half-skyrmion phase, see Appendix of \cite{PKLMR}.} 
 
 %%%%To continue %%%%
 
 Now contrast this to the results of what I consider as an unmistakeable indication of ``hadron-quark crossover" at density $n_{HQ}\simeq n_{1/2}$.  What's given in Fig.~\ref{sound} (bottom) is the sound velocity worked out in \cite{holt}  in what I classified above as a CDF theory with no {\it explicit} account of a possible hadron-quark crossover/transition mechanism. The different (colored) curves represent the predictions for the input parameters of $M(M_\odot)$ vs. $R$ (middle).  Up to density $\sim 2.5 n_0$, the sound speeds are essentially what was understood in the classic nuclear physics prediction~\cite{panda}, but increase continuously above the conformal velocity $v_c^2/c^2=1/3$ with no resemblance to the ``bump" at $\sim 2n_0$ seen in G$n$EFT.\footnote{In fact current detailed analyses fill up  the whole space for $n\gsim 2n_0$, both bottom-up and top-down. The bump feature of $v_{pcs}$ caused by an interplay of hadron-quark degrees of freedom, if present, is buried in the complex panorama. }

Although there are large number of papers in the literature dealing with the sound velocity in various models and with extensive analyses of experimental observations,  as far as I can see, at present there are no results whatsoever  that can rule out, theoretically or observationally,  the  simple structure  Fig.~\ref{sound} (top) predicted in G$n$EFT.  As it stands, the current situation is a jungle, much too complicated and confusing to give arguments to either support or rule out the ``simple a structure" $v_{pcs}$ at the Fermi-liquid fixed point level.  This is also the case in the calculations done top-down from asymptotic QCD.\footnote{There  appear in the literature results of  ``tour-de-force" {\it ab initio} QCD calculations from various groups of theorists, notably from Helsinki but  the results are much too unwieldy as they stand.}
An important point to make here is the sound velocity predicted in G$n$EFT is {\it not} conformal inside the massive compact stars although $v_{pcs}^2/c^2\sim 1/3$
resembles the conformal speed. In fact conformal symmetry is not restored with $v_{cs}^2/c^2=1/3$ in the range of density plotted in Fig.~\ref{sound} (top): The $\theta^\mu_\mu$ is not equal -- although rather  close -- zero in the center of the stars giving
\be
\Delta=\frac{1}{3} - \frac{P}{\epsilon} > 0
\ee
but it is density-independent, making its derivative go to zero. $\Delta=0$ is indeed the signal for conformal invariance but $v_{pcs}^2/c^2=1/3$ is not. What is found in  G$n$EFT 
is that not until the density for the DLFP is reached can the dense matter turn truly conformal.  The center of compact stars is highly unlikely to reach it. This is the reason why the sound velocity predicted in the central density of the compact star $\sim (6-7)n_0$ is referred to as ``pseudo-conformal velocity." It should however be noted that the polytropic index $\gamma$
\be
\gamma=d{\rm ln P}/d{\rm ln}\epsilon\label{pindex}
\ee
 plotted in Fig.~{\ref{comp} (right panel) behaves similarly to what could happen with the  ``deconfined" quarks in the conformal phase~\cite{annala}.
\begin{figure}[h]
%\begin{figure}
\begin{center}
\includegraphics[width=3.6cm]{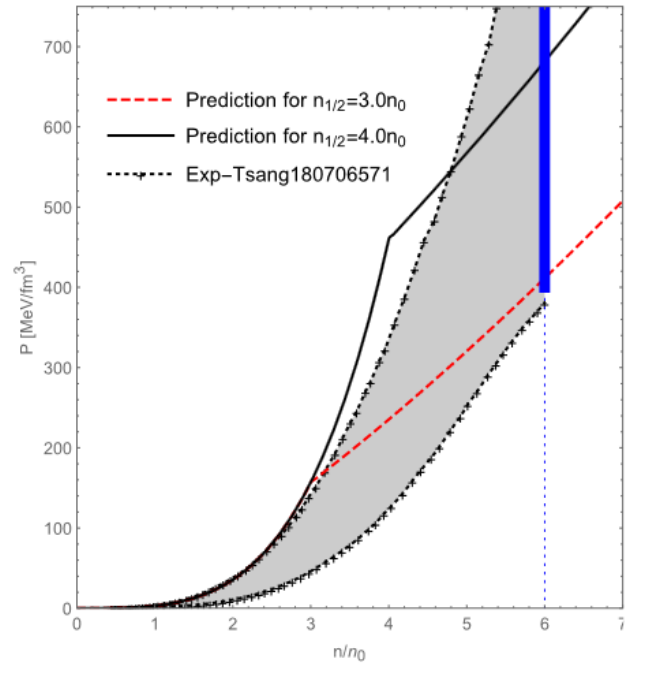}\includegraphics[width=0.3\textwidth]{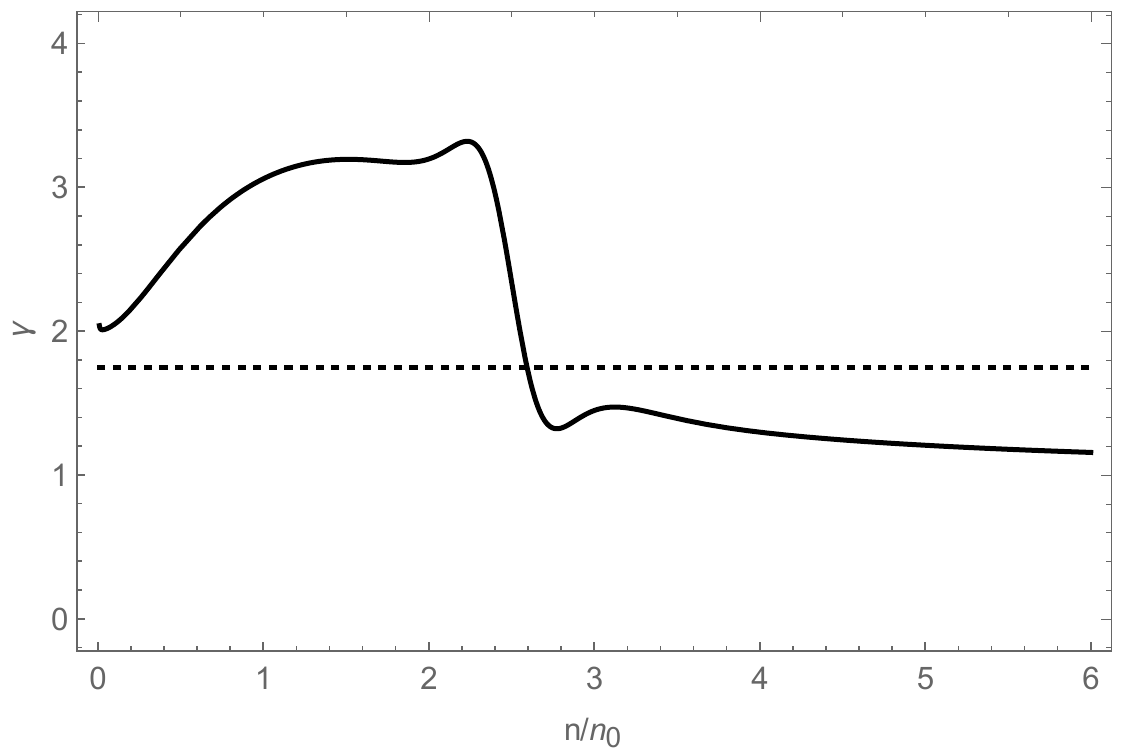}
\end{center}
\vskip -.5cm
\caption{Left panel: For $n_{1/2}=4n_0$,  the unitarity bound $v_s^2/c^2=1$ is violated.  A rough estimate gives the maximum star mass $M_{\rm max} \sim 2.3 M_\odot$.   Right panel:   Density dependence of  the polytropic index  in neutron matter for $n_{1/2}\approx 2.5 n_0$}
\label{comp}
% \end{center}
\end{figure}

One may be able to understand this difference between conformality and pseudo-conformality in the following way. In some profound way not yet fully clarified, it is again the interplay between the $\omega$ and the scalar $\sigma$ taking place in Walecka's linear mean-field model at low  density that continues effective  beyond $n\geq n_{HQ}$,  leading to the {\it emergent baryon parity-doubling} which is absent in the Lagrangian (\ref{Lag}), that is,  in the gauge theoretic QCD. This makes $\theta^\mu_\mu$ a constant $\la\chi_d\ra^4\propto m_0^4 \neq 0$ where $m_0$ is a chiral scalar, thus arriving at the pseudo-conformality~\cite{MaRho-Rev}. How this feature could manifest in the top-down scenario in QCD where quarks transform to hadrons is a challenging question.  

The possibility of this scenario in G$n$EFT is detailed (though with caveats) in \cite{PKLMR,MaRho-Rev} but there is one serious difficulty mentioned in these references that remains unresolved up to date. That has to do with the quantum (trace) anomaly, namely the  role of the anomalous dimension $\beta^\prime$. This is an important point which is somewhat obscured in the role  played by the hadron-quark crossover  when phrased in terms of the skyrmion-1/2-skyrmion transition. In fact, the qualitative structure of the EoS given by the Lagrangian (\ref{Lag}) at high density can actually  be gotten by dropping  entirely the baryon fields and looking at skyrmions on the  crystal lattice as density increases,  which becomes more reliable at higher density. In this approach,  one of the homogeneous Wess-Zumino (hWZ) terms in the HLS Lagrangian~\cite{HLS} --  non-topological for $N_f\neq 3$ --  which plays no significant role if  baryon fields are explicitly present, is  the only source for  the $\omega$ fields in the EoS at increasing density and brings in strong repulsion between the skyrmions (as baryons). That term is scale-invariant in the action, so the  dilaton attraction cannot counter-balance the repulsion.  This makes the  EoS of the skyrmion matter diverge.  This divergence can, however, be prevented if the dilaton comes in via an anomalous dimension $\beta^\prime \gsim 2$.~\cite{BRV}. How to translate this ``disaster" into the property of  G$n$EFT which has to do with the $\omega-\sigma$ interplay is yet to be clarified. I will point out below another puzzle involving the same $\omega-\sigma$ interplay and the pseudo-conformality in connection with quenched $g_A$s in nuclear Gamow-Teller transitions.

% coadjoint orbits.   $\cal{L}_{\rm GnEFT}$,

To conclude, I make  a brief summary -- at the risk of redundancy -- of  the relays that go into the review of G$n$EFT and possible open issues remaining unresolved. 

First the mean-field structure of the $\omega-\sigma$ linear model~\cite{walecka} is extended to a multi-field structure bringing in corrections by higher-mass fields which remedy the defects of the linear model.   This step leads to what may be considered as an improved CDF-type theory.  By fitting  parameters to available experimental data, one arrives at a fine-tuned CDF such as is the one worked out  in \cite{holt}. One might call it a phenomenological CDF. Such a CDF could be matched to low-order  chiral Lagrangian anchored on Weinberg's ``Folk Theorem"   on EFT, exemplified by $\chi$EFT$_\pi$, recognized in particle theory~\cite{gelmini-ritzi}  and later in nuclear theory~\cite{DD}.  Put on a Fermi sphere, a Landau(-Migdal) Fermi liquid-type structure  enables one to address  compact stars~\cite{matsui}. By formulating in renormalization-group (RG) approach strongly  interacting fermions on Fermi surface, one arrives in the high density limit at an EFT given at the Fermi-liquid fixed-points~\cite{shankar,polchinski}. Suitably implemented with relevant massive hidden symmetry degrees of freedom (hidden local symmetry,  hidden scale symmetry, parity doubling symmetry etc.) combined with a topology change accounting for hadron-quark continuity at $n_{HQ} \sim 2.5n_0$, one then arrives at an improved CDF,   G$n$EFT, that can be applied for the {\it whole} range of density from $n_0$ to dense compact-star matter density ${\sim  (6-7)n_0}$~\cite{PKLMR}. One may therefore consider G$n$EFT a ``Generalized Conformal Density Functional." 
Now incorporating couplings between the strongly correlated baryons on the Fermi surface and the fluctuating hidden symmetry degrees of freedom in the coadjoint-point  technique~\cite{coadjoint} being developed in condense matter physics, one could arrive at systematic corrections to the Fermi-liquid fixed-point results that go beyond the $1/\bar{N}$ expansion~\cite{shankar,polchinski}.  It would be interesting to see whether such corrections could bring deviations from the ``simple" structure of $v_{psc}$ given in the FLFP approximation.

In G$n$EFT where the Lagrangian $\cal{L}_{\rm GnEFT}$ is treated in the FLFP approximation,  the EoS, such as $E_{sym}$ and its derivatives,  do come out consistently with  the treatment of \cite{holt} up to $\sim n_{HQ}\approx n_{1/2}$ density. But for $n > n_{HQ}$, there must be differences. Indeed one expects this difference in the sound speed in the center of massive stars. 

As a whole, apart from the sound speed, which might ultimately get into conflict with future observable constraints with precision, so far the G$n$EFT predictions~\cite{PKLMR, MaRho-Rev} face no serious conflicts with the available observables, such as $M^{max}$, $R_{2.0}$, $R_{1.44}$ etc.\footnote{There seems to be some tension with the predicted tidal deformability $\Lambda_{1.4}\sim 550$ with the observation from GW1700817.  But  the theoretical prediction depends  very sensitively on the precise value of $n_{HQ}$ which cannot be precisely pinned down.  So the predicted value cannot be taken too seriously.}

Let me finally return to the puzzling  -- and possibly fundamental--  issue mentioned above re: the intricate interplay of the $\omega$ and $\sigma$ (or $\sigma_d$)  in connection with  the anomalous dimension $\beta^\prime_{\rm IR}$ in nuclear $\beta$ decay. Leaving details to \cite{2-neutrino},  the currently accepted  axial current coupling constant effective in nuclear processes is $g_A^{\rm eff}\approx  1$.  It is  found however that the effective $g_A$ in the most recent Gamow-Teller transition in the doubly magic-shell nucleus $^{100}$Sn measured at RIKEN considered to be most reliable gives a substantially bigger quenching,  $g_A^{\rm eff}\sim (0.6-0.8$).  Such a quenching can seriously impact the search for the BSM given that the effective axial constant figures as $\sim (g_A^{\rm eff})^4$.  It turns out that the ``disaster" could be avoided if the anomalous dimension were $\beta^\prime \gsim 2$~\cite{BRV} whereas in the current development of scale symmetry~\cite{zwicky} adopted for G$n$EFT, it is predicted $\beta^\prime \sim 0$. 

\subsection*{I would like to acknowledge sharing the initial ideas with Hyun Kyu Lee and Won-Gi Paeng on the role of the ``hidden" interplay of the $\omega$  and the dilaton $\sigma_d$ in highly dense matter.}

 \end{document}